\documentclass{article}
\pdfoutput=1
\usepackage[utf8]{inputenc}
\usepackage[margin=1in]{geometry}
\usepackage{setspace}
\usepackage[colorlinks=true]{hyperref}
\hypersetup{
  colorlinks = true,
  citecolor = blue,
  urlcolor = blue,
  linkcolor = blue,
}
\usepackage{amsmath}
\usepackage{amssymb}
\usepackage{graphicx}
\usepackage{natbib}
\setcitestyle{numbers,square}
\usepackage[font=footnotesize]{caption}
\usepackage{afterpage}

\title{Wave function representation of probability distributions}
\author{Madeleine B.\ Thompson}
\date{2018--01--05}

\begin{document}
\maketitle

\begin{abstract}
Orthogonal decomposition of the square root of a probability density function
in the Hermite basis is a useful low-dimensional parameterization of continuous
probability distributions over the reals.  This representation is formally
similar to the representation of quantum mechanical states as wave functions,
whose squared modulus is a probability density.
\end{abstract}

\section{Motivation}

Empirical Data Platform (EDP) is a cloud platform for data analysis. For
modeling regression residuals, it requires a simple, natural, low-dimensional
family of smooth probability distributions over the reals. This family must be
able to represent a wide variety of real-world data without tuning or
customization.

With one free parameter, the natural distribution is a Dirac delta. With two
free parameters, the natural distribution is a Gaussian.  With three or more
free parameters, I know of no consensus.  The Pearson
distribution\cite[\S20, p.~381]{pearson} is one general family,
but it does not have a clean parameterization. Johnson's $S_U$
distribution\cite{johnson} and the generalized lambda
distribution\cite{ramberg} are both four-parameter families, which have more
flexibility than normal distributions do but don't support multimodal
distributions, which are necessary for modeling residuals in EDP.
Fleishman\cite{fleishman} uses a four-term polynomial sum of Gaussians, which
doesn't allow for asymmetric distributions.  Gentle\citep[p.~193--196]{gentle}
has further discussion.

So, we still require a class of distributions that can handle skewed data with
several modes, preferably without the complexity and computational cost that
come with nonparametric methods like kernel density estimation.

\section{Details}

Suppose we have a random variable $X$ with a density function $f(x)$. Assume it
has been transformed so that its mean is approximately zero and its variance is
approximately one-half.

\begin{figure}
\includegraphics{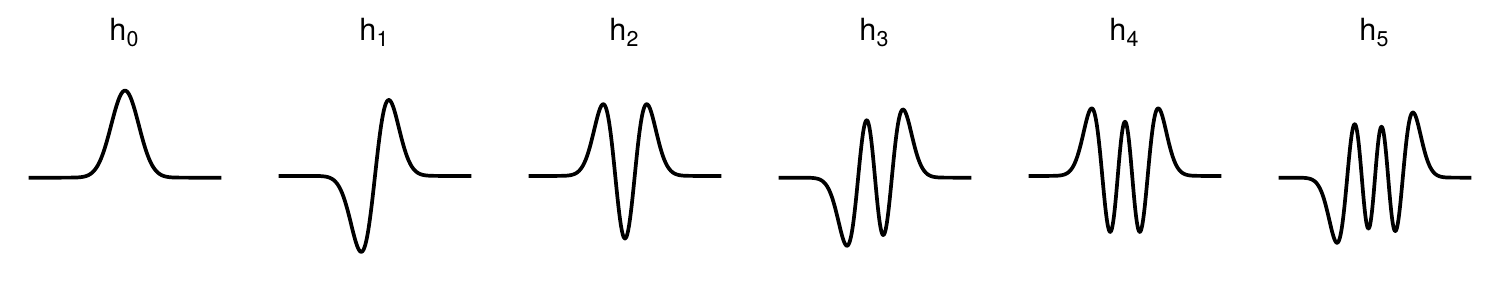}
\caption{The first several basis functions defined by equation~\ref{hn}.  Only
$h_0$ is always positive, so while these can be used as a basis for the
square root of a density, they can't be used as the basis for an untransformed
density.}
\label{basis}
\end{figure}

Let's find a function $\phi(x)$ such that $f(x) = \phi(x)^2$, where $\phi(x)$
is a sum of orthogonal components.  In a quantum mechanics context, $\phi(x)$
would be an amplitude. Here, we use the Hermite basis, since it makes Gaussians
come out nicely. Let $H_n$ be the physicists' Hermite
polynomial of degree $n$.\cite[table~22.12, p.~801]{abramowitz} Define $h_n(x)$
as:
\begin{equation}\label{hn}
h_n(x) = H_n(x) \frac{e^{-x^2/2}}{\left(\sqrt{\pi} 2^n n!\right)^{1/2}}
\end{equation}
The first several of these are shown in figure~\ref{basis}. This normalization
of the Hermite polynomials is orthonormal:
\begin{equation}
\int_{-\infty}^\infty h_n(x) \, h_m(x) \, dx = \delta_{nm}
\end{equation}
Pick a maximum degree $K$. (EDP defaults to $K=10$.) Define a vector $w \in
\mathbb{R}^{K+1}$ by:
\begin{align}\label{wk}
w_k &= \int_{-\infty}^\infty h_k(x) \, \sqrt{f(x)} \, dx \\
\text{where } k &\in \{0, \ldots, K\}
\end{align}
Observe that the square root of the density for $N(0, \frac12)$ is:
\begin{align}
\sqrt{f(x)} &= \sqrt{\frac1{\sqrt{2\pi (1/2)}} e^{-\frac{x^2}{(2)(1/2)}}} \\
&= \frac1{\sqrt{\sqrt{\pi}}} e^{-x^2/2} \\
&= h_0(x)
\end{align}
So, $N(0, \frac12)$ is represented exactly by $w = (1, 0, 0, \ldots, 0)$.

Next, define a degree-$K$ approximation to $f$:
\begin{equation}\label{fhat}
\hat f(x) = \left( \sum_{k=0}^K w_k \, h_k(x) \right)^2
\end{equation}
$f$ is a density, so it must be in $\mathcal{L}^1$. Therefore, $\sqrt{f} \in
\mathcal{L}^2$. Since the Hermite basis is complete for $\mathcal{L}^2$, $\hat
f$ converges in $\mathcal{L}^1$ to $f$ as $K \rightarrow \infty$. If $f$ is
symmetric, then $w_k$ is zero for odd $k$. The sum $\sum_{k=0}^K w_k^2$
converges to one as $k\rightarrow\infty$; the partial sum can be used as a
check of how well $\hat f$ matches $f$.

Now consider the quantum harmonic oscillator with potential $V(x) \propto x^2$.
The coefficients $w$ are the amplitudes in the Hamiltonian basis for the state
whose initial position is $f(x)$.\cite[p.~56]{griffiths}

\section{Estimating the coefficients}\label{estimating}

The formula for $w$, equation~\ref{wk}, contains an integral. If we knew $f$,
it would be easy to compute this numerically using adaptive Simpson's
method\cite[ch.~6]{miller} or Gauss--Hermite
quadrature.\cite[eqn.~25.4.46]{abramowitz}.

However, we would like to model regression residuals, so we need to be able to
fit $\{w_k\}$ from an i.i.d.~sample. We can do this by computing the MLE of $w$
on a transformed space. Consider the unit $K$-sphere in $\mathbb{R}^{K+1}$
(noting the zero-based indices):
\begin{equation}
w_0^2 + \cdots + w_K^2 = 1
\end{equation}
This is the set of admissible $w$. Outside this sphere, the probability density
$\hat f$ would integrate to more than one. For any $w$ on the surface of the
sphere, consider the line through $(1, 0, \ldots, 0)$ and $w$. Define $P: w
\mapsto \gamma$ to be the mapping from $w$ to the point on that line where the
line intersects the plane whose first coordinate is zero. This is a
stereographic projection of $w$.\cite[ch.~2]{spivak} By dropping
the coordinate with index 0 (which is always zero), we have transformed $w$,
which is constrained to be on the surface of a sphere, to $\gamma$, which has
one fewer dimension but is unconstrained. Every value on this plane except for
the origin maps to a unique $w$ on the unit sphere. The pre-image of the origin
could be considered to be either $(1, 0, \ldots)$ or $(-1, 0, \ldots)$. I choose
$(-1, 0, \ldots)$ for continuity, though they both lead to the same probability
density.

This is the algebraic form of the transform\cite{stereo}:
\begin{align}
[P(w)]_k &= \frac{w_k}{1-w_0} \\
k &\in \{1, \ldots, K\}
\end{align}
This is its inverse:
\begin{align}\label{invgamma}
[P^{-1}(\gamma)]_k &= \begin{cases}
\frac{S^2-1}{S^2+1} & \text{if } k = 0 \\
\frac{2\gamma_k}{S^2+1} & \text{if } 1 \leq k \leq K
\end{cases} \\
\text{where } S^2 &= \sum_{k=1}^K \gamma_k^2 \\
\text{and } k &\in \{0, \ldots, K\}
\end{align}

With these definitions, the log likelihood at $\{x_1, \ldots, x_n\}$ is:
\begin{equation}
\ell(\gamma) = \sum_{i=1}^n \log\left( \sum_{k=0}^K [P^{-1}(\gamma)]_k \cdot h_k(x_i) \right)^2
\end{equation}
I have found L-BFGS\cite{liu}, as implemented by libLBFGS\cite{okazaki}, to work
well for finding the MLE of $w$, even for sample sizes as small as two.  I
expect that Gibbs sampling or Hamiltonian Monte Carlo\cite{nealhmc} would work
well for drawing $w$ from a Bayesian posterior.

\section{Moments, entropy, and sampling}

Because the density $\hat f$ is a polynomial times $\exp(-x^2)$, Gauss--Hermite
quadrature\cite[eqn.~25.4.46]{abramowitz} can be used to exactly (up to
floating point roundoff) compute any moment in $O(K)$ primitive floating point
operations (no expensive transcendental functions).  Non-polynomial integrands,
such as the entropy integral $-\int f(x) \log f(x) \,dx$, are not exact when
computed this way, but this method is still more accurate than Monte Carlo
estimation for a comparable amount of computation time.

EDP is currently using a univariate slice sampler\cite{nealslice} to sample
from $\hat f$.  Since we normalize $\{x_i\}$ to have variance one-half before
fitting $\hat f$, EDP can use a fixed initial slice width, 4.0.  Because the
derivative of the density estimate (equation~\ref{fhat}) is simple, I plan to
switch to Adaptive Rejection Metropolis Sampling (ARMS)\cite{gilks} if EDP ever
needs a faster sampler.

\section{Examples}

\begin{figure}
\includegraphics{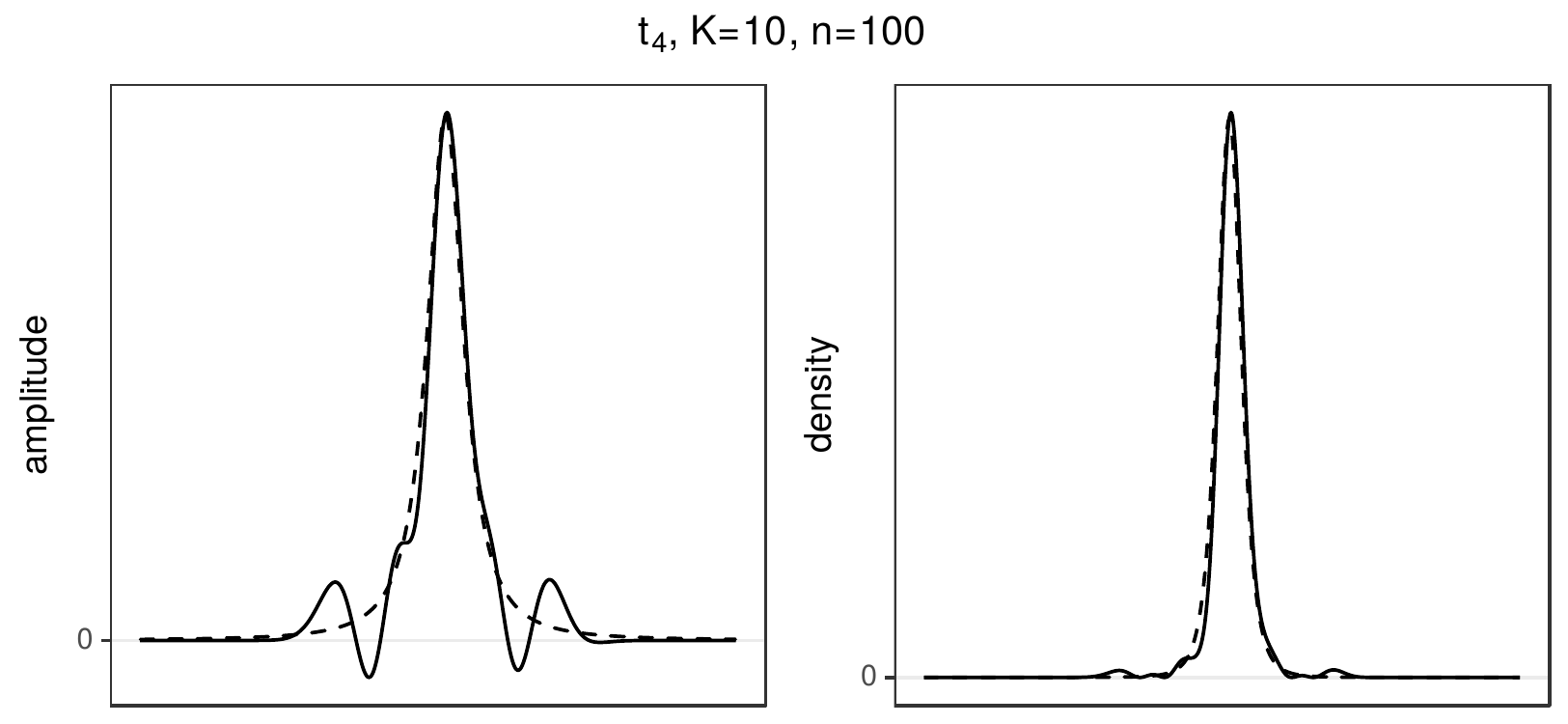}
\caption{This Student $t$ with four degrees of freedom does not have moments
greater than 3, but there is no issue fitting the tenth-degree Hermite
expansion. Observe that since the MLE optimizes fit in \textit{squared}
amplitudes, most discrepancy between $\sqrt{f(x)}$ and the amplitude estimate
in the left plot is in places where both are close to zero.}
\label{firstex}
\end{figure}

\begin{figure}
\includegraphics{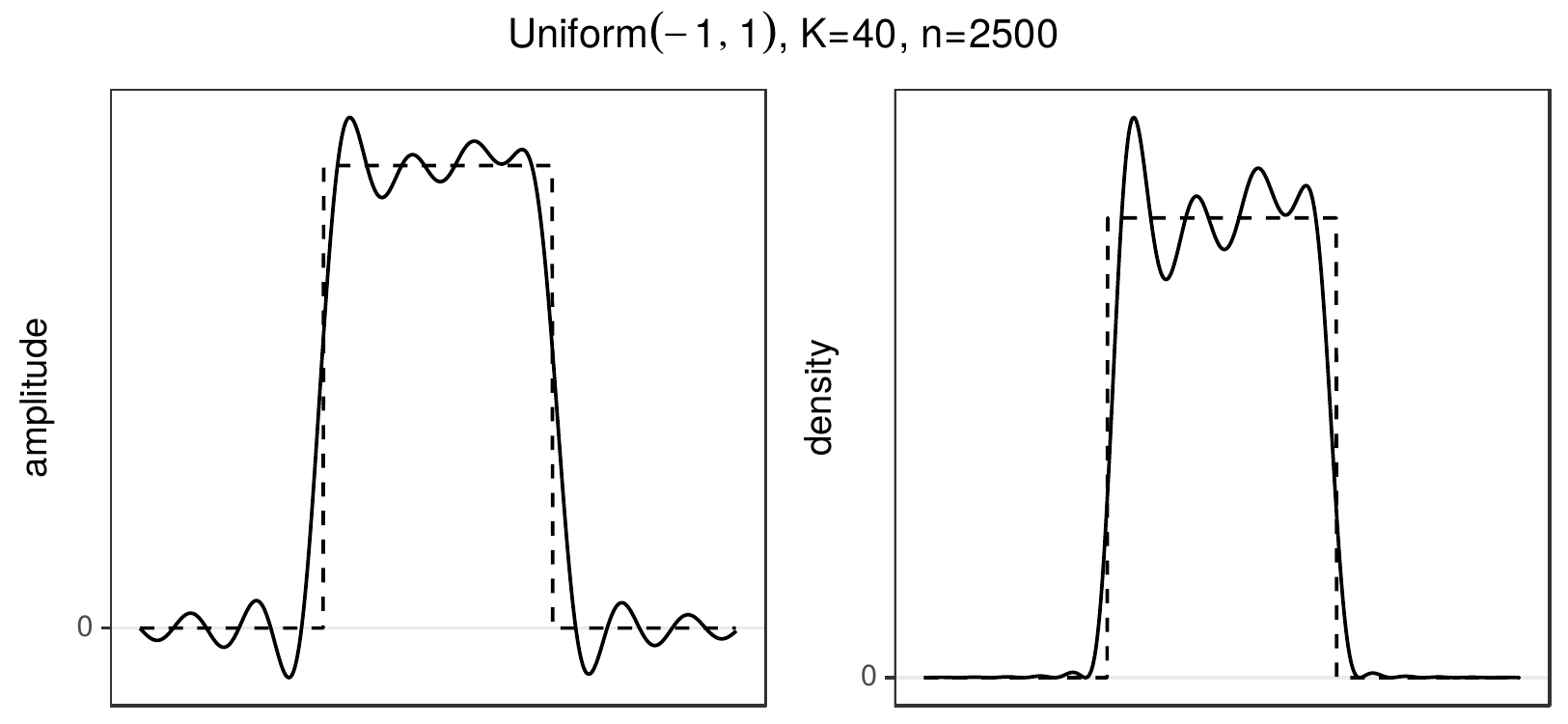}
\caption{The uniform distribution is difficult to fit because its density is
discontinuous. Notice that even with a degree-40 representation and a
sample size of 2500, the fit is not good. We observe the Gibbs phenomenon, just
as we would in the more familiar case of a Fourier expansion of a square wave.
The fit does improve with increasing degree (in the sense of total variation
distance), but the normalization constant in $h_k$ (equation~\ref{hn}) is
$O(\sqrt{2^k k!})$, which causes floating point errors for $K$ larger than
around 20.}
\label{bigunif}
\end{figure}

\begin{figure}
\includegraphics{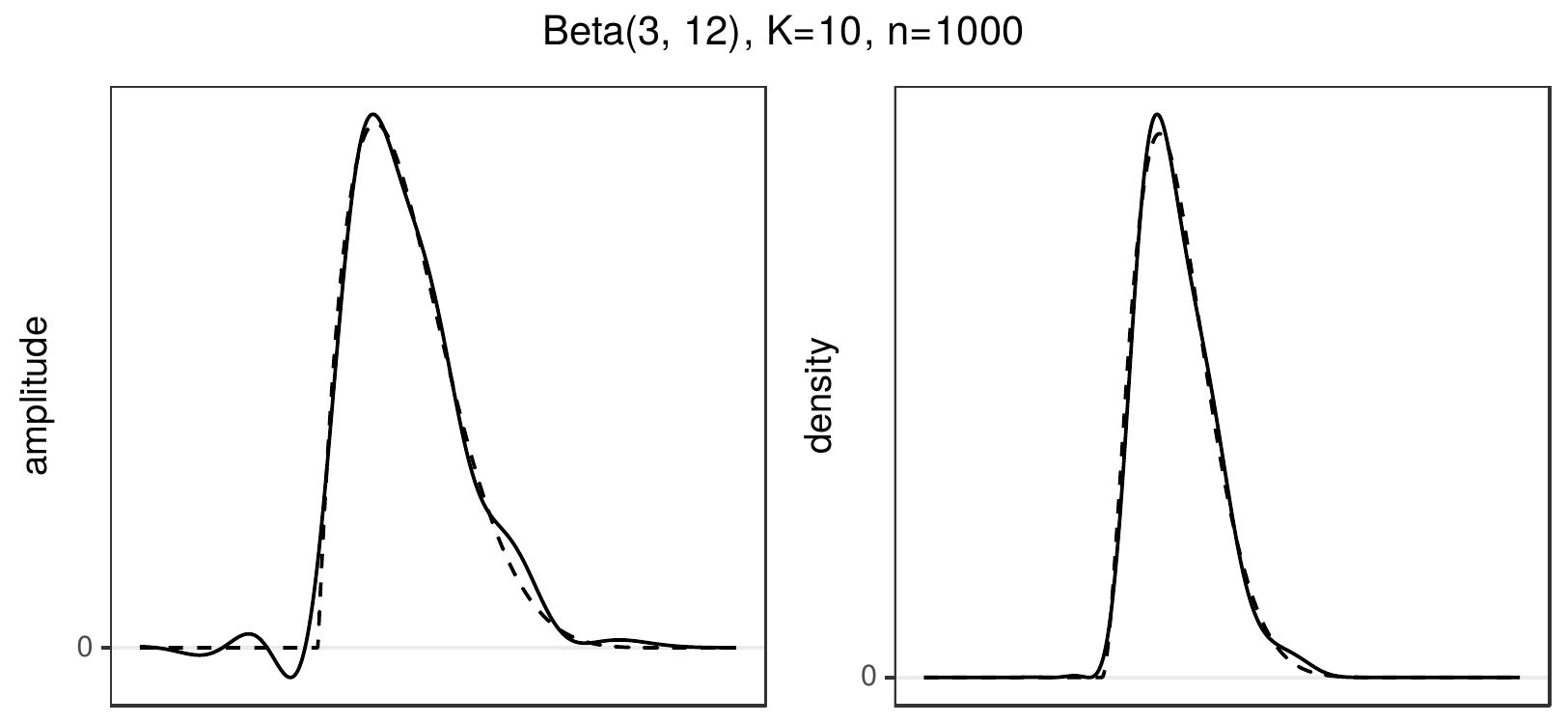}
\caption{This $\operatorname{Beta}(3, 5)$ distribution shown here demonstrates
that the wave function method can fit skewed distributions.}
\label{lastex}
\end{figure}

Figures~\ref{firstex} through \ref{lastex} are examples of distributions I used
for testing the match between the true density and the wave function
representation. In each figure, the left plot shows the amplitude, and the
right plot shows its square, the density. The dashed lines are the true
amplitude $\sqrt{f(x)}$ and density $f(x)$, and the solid lines are the
amplitude estimate $\sum w_k h_k(x)$ and the density estimate $\hat f(x)$
obtained by computing the MLE of $w$.

\begin{figure}
\includegraphics{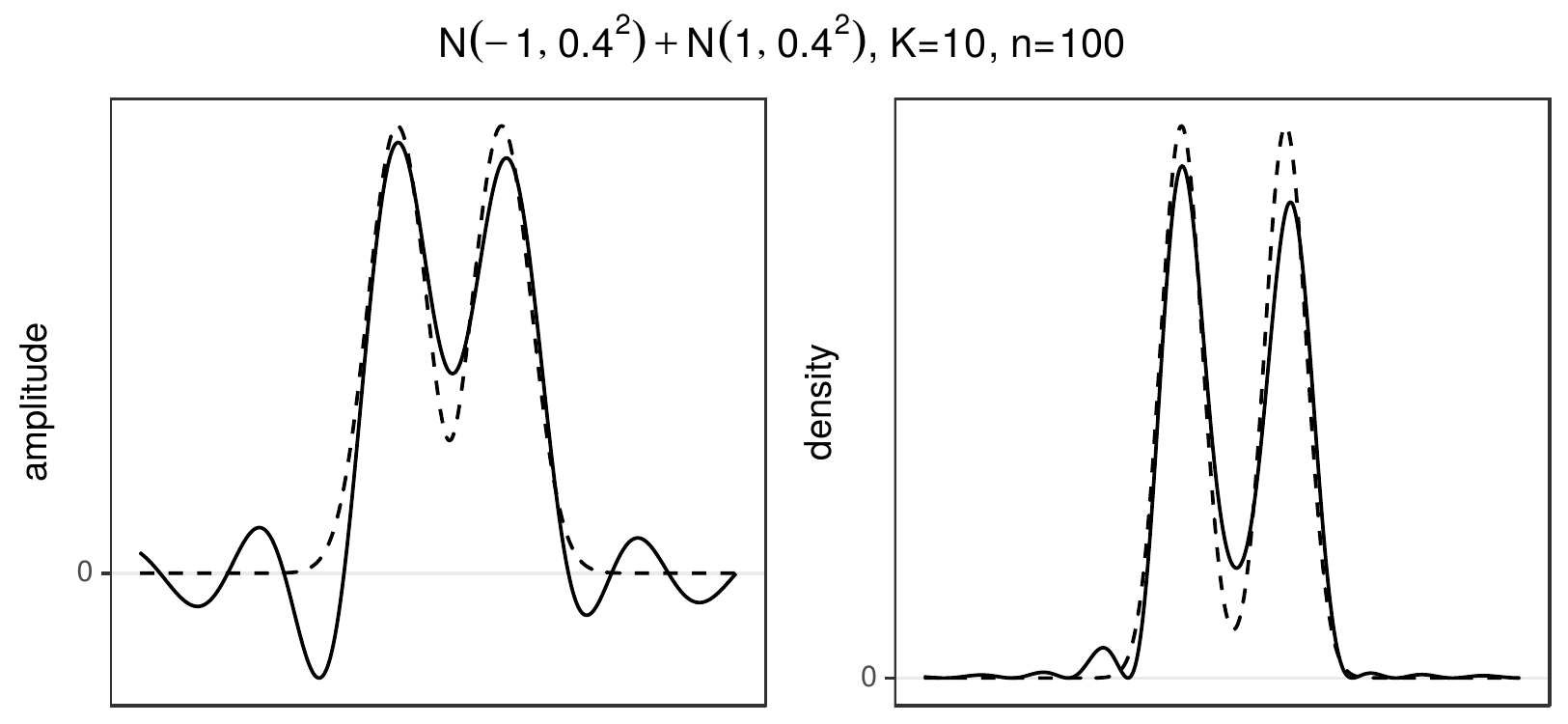}
\caption{This plot of a bimodal normal mixture shows the limitations of
low-degree polynomials in representing separated modes. Raising the degree from
10 to 12 or 14 substantially improves the fit. Larger sample sizes fail to make
a large difference, except to pin down the exact relative heights of the
modes.}
\end{figure}

\section{Discussion}

The wave function representation of continuous probability densities is a
practical solution to the need for a general class of well-behaved probability
densities. It can represent any smooth density, yet is resistant to
over-fitting. Unlike, for example, kernel density estimation, it involves no
tuning parameters. Also, unlike kernel density estimation, it has nicely shaped
tails. Coefficients can be fit quickly with off-the-shelf methods.  As a
result, the wave function representation has been effective in a production
data analysis system for modeling a wide variety of user-uploaded data.

This paper discusses only unconditional densities with support on the real
line. The extension to densities on different spaces, using different bases, is
straightforward. For example, the Legendre polynomials\cite[ch.~8,
p.~333]{abramowitz} could be used in place of the Hermite polynomials for
modeling distributions on finite closed intervals of the real line. Relatedly,
I wonder whether the solutions to Schrödinger's equation for simple potentials
other than $x^2$ could yield classes of probability distributions useful
outside quantum physics. Finally, I hope to explore the possibility of fitting
conditional models, where the coefficients $w$ are functions of a predictor,
for heteroskedastic regressions.

R code for fitting wave functions to distributions is available in the
``wavefunction'' package on CRAN.\cite{rpkg}

\afterpage{\clearpage}

\bibliographystyle{plainurl}
\bibliography{references.bib}

\end{document}